\documentclass[conference]{IEEEtran}
\IEEEoverridecommandlockouts

\usepackage{cite}
\usepackage{amsmath,amssymb,amsfonts}
\usepackage{graphicx}
\usepackage{textcomp}
\usepackage{xcolor}
\usepackage{quantikz}
\usepackage{caption}
\usepackage{soul}
\usepackage{xspace}
\usepackage{float}
\usepackage{multirow}
\usepackage{grffile}
\usepackage{subcaption}
\usepackage{makecell}
\usepackage{amsmath,amsfonts}
\usepackage{algorithm}
\usepackage[noend]{algorithmic}
\usepackage{amssymb}
\usepackage{comment}
\usepackage{url}

\newcommand{\sol}{CCMap}

\def\BibTeX{{\rm B\kern-.05em{\sc i\kern-.025em b}\kern-.08em
    T\kern-.1667em\lower.7ex\hbox{E}\kern-.125emX}}
\begin{document}

\title{Hardware-aware Compilation for Chip-to-Chip Coupler-Connected Modular Quantum Systems}

\author{
\IEEEauthorblockN{
Zefan Du\IEEEauthorrefmark{1},
Shuwen Kan\IEEEauthorrefmark{1},
Samuel A. Stein\IEEEauthorrefmark{2},
Zhiding Liang\IEEEauthorrefmark{3},
Ang Li\IEEEauthorrefmark{2},
Ying Mao\IEEEauthorrefmark{1}
}
\IEEEauthorblockA{\IEEEauthorrefmark{1}Department of Computer and Information Sciences, Fordham University, NY, USA\\
Email: \{zdu19, sk107, ymao41\}@fordham.edu}

\IEEEauthorblockA{\IEEEauthorrefmark{2}High Performance Computing Group, Pacific Northwest National Laboratory, WA, USA\\
Email: \{samuel.stein, ang.li\}@pnnl.gov}

\IEEEauthorblockA{\IEEEauthorrefmark{3}Department of Electrical, Computer, and Systems Engineering, Rensselaer Polytechnic Institute, NY, USA\\
Email: liangz9@rpi.edu}
}

\maketitle

\begin{abstract}
As quantum processors scale, monolithic architectures face growing challenges due to limited qubit density, heterogeneous error profiles, and restricted connectivity. Modular quantum systems, enabled by chip-to-chip coupler-connected modular architectures, provide a scalable alternative. However, existing quantum compilers fail to accommodate this new architecture. 
We introduce CCMap, a circuit-compiler co-design framework that enhances existing quantum compilers with system-level coordination across modular chips. It leverages calibration data and introduces a coupler-aligned and noise-aware cost metric to evaluate circuit compilation. CCMap integrates with existing compilers by partitioning circuits into subcircuits compiled on individual chips, followed by a global mapping step to minimize the total cost. We evaluated CCMap on IBM-Q noisy emulators using real hardware calibrations across various coupler-connected topologies. Results show that CCMap improves circuit fidelity by up to 21.9\%, representing a 30\% increase, and reduces compilation cost by up to 58.6\% over state-of-the-art baselines. These findings highlight CCMap’s potential to enable scalable, high-fidelity execution in coupler-connected modular quantum systems.

\end{abstract}

\section{Introduction}

Quantum computers provide a transformative approach for enhancing existing use cases and building new applications~\cite{li2025quantum, l2024quantum, d2023distributed, ruan2023quantumeyes, stein2021qugan}. Achieving utility-level quantum advantage likely demands hardware with many more physical qubits than currently available hardware, as quantum error correction relies on redundancy to construct reliable logical qubit~\cite{gidney2021factor}. Scaling quantum computation requires both hardware and software advancement.

Despite recent progress from IBM, Google, IonQ, and Quantinuum including IBM's 1121-qubit Condor, monolithic superconducting systems still face critical scalability challenges of large-scale quantum hardware. These include qubit decoherence, control complexity, cooling limitations, and especially fabrication-induced defects that degrade overall fidelity as chip sizes increase~\cite{mohseni2025how, ibm2024tls}

\begin{figure}[htbp]
    \centering
    \includegraphics[width=1\linewidth]{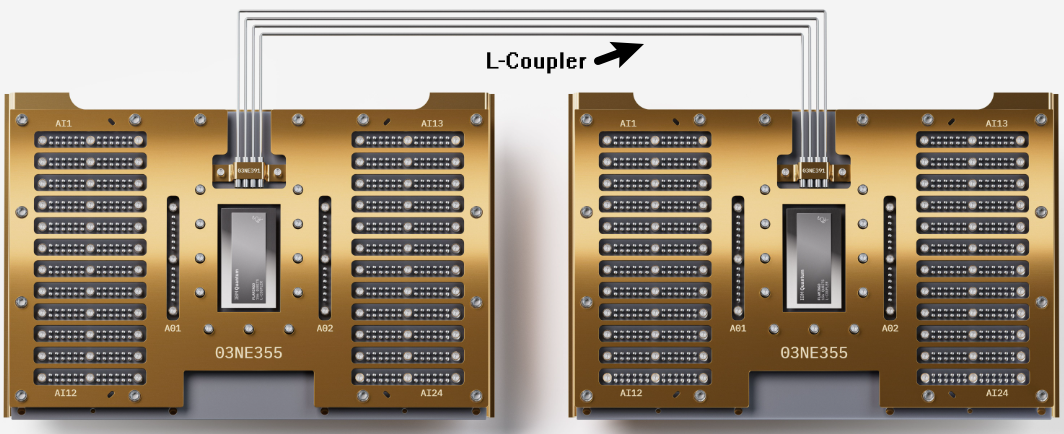}
    \caption{IBM Flamingo Architecture~\cite{ibmQuantumDelivers}}
    \label{fig:flamingo}
\end{figure}

To address scaling challenges, modular quantum architectures offer a path forward. In a modular quantum system, multiple smaller processors are connected by on-board couplers that provide fixed links for executing larger circuits~\cite{ChipletScalingSmith}. Figure~\ref{fig:flamingo} shows IBM’s Flamingo architecture, which connects Heron R2 chips using meter-long L-Couplers~\cite{ibmQuantumDelivers}, enables direct inter-chip operations and extends capacities beyond the limitations of monolithic devices.

For a circuit  compilation perspective, however, modular systems introduce new challenges that traditional compilers are not equipped to handle. These include heterogeneous noise profiles, fragmented topologies with limited coupler links, and dynamic fidelity variations across chips. Recent works have also highlighted how hybrid optimization strategies~\cite{10821429} and circuit cutting in distributed systems~\cite{du2024efficientcircuitcuttingscheduling} require compilers to adapt to architectural constraints. Furthermore, resource-aware quantum execution and scheduling in hybrid systems~\cite{jiang2024resourceefficientselfadaptivequantumsearch} and fault-tolerant compilation approaches targeting surface-code Pauli-based optimization~\cite{kan2025sparosurfacecodepaulibasedarchitectural} emphasize the need for hardware-adaptive strategies.

Existing quantum compilers are designed for single-chip devices and focus on monolithic SWAP optimization~\cite{sabre2019, ibm_qiskit} and assume uniform noise models~\cite{cowtan_et_al_pyket}, lacking support for fragmented topologies, and failing to exploit the distinct characteristics of coupler-connected systems~\cite{murali2019noise, wagner2025optimized}. Effective compilation for modular quantum systems must co-optimize local qubit mappings with inter-chip routing, accounting for the coupler-induced noise and topological fragmentation.

In this work, we introduce \sol, a lightweight circuit-compiler co-design framework. Specifically, CCMap works on top of existing compilers and boost their performance in modular quantum systems. 
\sol~ coordinates compilation across multiple chips by analyzing circuit structure and qubit interactions to resolve inter-chip dependencies and minimize a coupler-aligned and noise-aware cost function that reflects both on-chip and coupler-induced errors. This enables effective transpilation for chip-to-chip coupler-connected modular quantum systems. Our key contributions are as follows: 
\begin{itemize}
    \item \textbf{Coupler-aligned and Noise-Aware Cost Modeling:} We propose a fidelity-oriented cost function that captures gate error rates and execution latencies for both on-chip and coupler links.

    \item \textbf{Topology-Aware Partitioning:} We partition the circuit’s entanglement graph to confine most two-qubit gates within chips and reduce inter-chip operations. Utilizing coupler links to connect subcircuits, CCMap eliminates costly post-processing reconstruction.

    \item \textbf{Seamless Compiler Integration:} \sol~integrates with existing compilers (e.g., Qiskit), enabling scalable compilation for large circuits on modular quantum systems.
    
\end{itemize}

Overall, this co-design approach provides a robust and extensible solution for coupler-aware compilation in monolithic compilers, targeting the specific challenges of chip-to-chip coupler-connected modular quantum systems.  With extensive experiments, results demonstrate that CCMap improves existing compilers by up to 21.9\% in fidelity and reduces compilation cost by up to 58.6\% over state-of-the-art baselines. \sol~offers a practical path toward executing large-scale quantum algorithms on modular hardware with future-scalable, hardware-aware transpilation strategies. 
\section{Related Work}

Quantum circuit compilation has been widely studied, particularly for monolithic devices. Early methods focused on depth and SWAP minimization, such as SABRE~\cite{sabre2019}, with later extensions introducing noise-awareness~\cite{lao2022}. Frameworks like MQT~\cite{mqtqmap} and UCC~\cite{ucc2025} support multi-objective optimization, while learning-based approaches predict qubit mappings from circuit structure~\cite{zhang2021}. However, these assume fixed-topology systems and lack support for chip-to-chip constraints like inter-chip noise and latency. Our work explicitly models such heterogeneity, assigning higher cost to coupler-mediated operations for fidelity-aware compilation.

To scale beyond monolithic designs, the most relevant recent work is the modular compiler by~\cite{jeng2025modularcompilationquantumchiplet}, which introduces a Stratify-Elaborate compiler architecture tailored for chiplet-based quantum systems. It does not incorporate circuit entanglement graph partitioning nor support external compiler integration, limiting its flexibility for heterogeneous or evolving toolchains. In contrast, our method is designed as a modular plugin that can interoperate with existing compilers.

In summary, while prior work tackles  noise-aware mapping~\cite{lao2022}, circuit partitioning~\cite{kan2024scalable}, and modular compilation~\cite{jeng2025modularcompilationquantumchiplet}, few approaches jointly optimize for coupler-aware noise, connectivity, and fidelity. Our work addresses this gap through co-designed partitioning and mapping tailored for coupler-connected modular systems.

\vspace{-0.05in}
\section{Background and Motivation}

This section outlines key concepts: chip-to-chip onboard coupler links, circuit graph partitioning, and noise-related challenges, all of which are fundamental to compilation in modular quantum systems with coupler-based interconnects.

\subsection{Modular Quantum Systems and Coupler Links}

Modular quantum architectures connect multiple chips using fixed coupler links, enabling larger capacity than monolithic devices. Inter-chip operations such as non-local $\sqrt{\text{SWAP}}$, $CX$ gates, and quantum state transfer (QST), expand circuit capacity and support distributed execution~\cite{ statetransfer2021}.

Couplers are implemented across platforms including superconducting microwave links~\cite{Magnard_2020}, optical interconnects~\cite{reiserer2015cavity}, and trapped-ion transport~\cite{ wang2020integrated}. Despite progress, these links suffer from higher error rates and noise than on-chip gates due to decoherence, signal loss, and imperfect entanglement generation~\cite{wang2020integrated}.

IBM's Flamingo architecture (Figure~\ref {fig:flamingo}) demonstrates inter-chip coupling between Heron R2 chips via meter-long superconducting links, achieving inter-chip $CNOT$ gate at a 3.5\% error rate~\cite{ibmQuantumDelivers}. However, further improvements in gate fidelities and noise mitigation remain essential for scalable distributed quantum computation.

\subsection{Circuit Partitioning}

Circuit cutting divides large quantum circuits into smaller subcircuits executable on near-term devices~\cite{peng2020simulating}. While wire and gate cuts reduce hardware requirements, reconstruction introduces exponential overhead. Recent approaches reduce cuts by analyzing entanglement and hardware constraints~\cite{tang2021, kan2024scalable, qiskit-cutting}.

This is especially important in modular architectures, where expensive inter-chip operations across coupler links should be minimized. By modeling both the circuit and hardware as coupling graphs, logical-to-physical qubit mapping becomes a graph partitioning task co-optimized to reduce inter-chip communication. Moreover, in modular quantum systems, utilizing coupler links to connect subcircuits can avoid costly classical reconstruction.

\subsection{Noises and Qubit Mapping}

Quantum states are highly sensitive to various noise from hardware, including decoherence, gate infidelities, and cross-talk~\cite{krantz2019quantum}. On-chip noise primarily arises from calibration errors in single-qubit gates~\cite{ballance2016high}, interaction imperfections in two-qubit gates~\cite{barends2014superconducting}, and electromagnetic cross-talk~\cite{mckay2017efficient}. Qubit mapping assigns logical to physical qubits—directly, impacting execution fidelity by minimizing noise and honoring hardware constraints~\cite{ibm_qiskit, tannu2019not}.

In modular quantum systems, inter-chip operations suffer from higher error rates due to entanglement generation variability and coupler-induced decoherence. This increases the complexity of qubit mapping, requiring strategies that minimize both on-chip and inter-chip noise. Recent methods use learning-based and graph-based techniques to adapt mappings dynamically using calibration data~\cite{murali2019noise}, improving overall fidelity in modular architectures.

\begin{figure}
    \centering
    \includegraphics[width=1\linewidth]{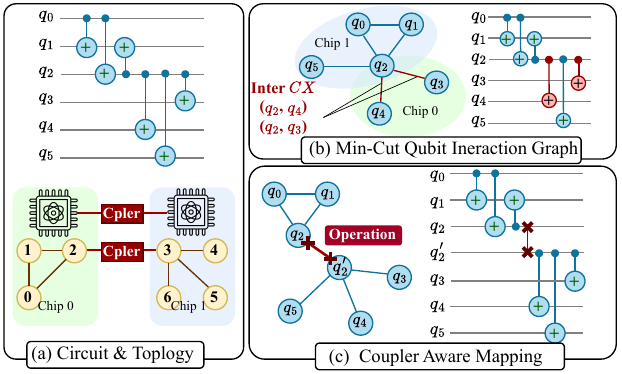}
    \vspace{-0.1in}
    \caption{Overview of the compilation flow for a coupler-connected modular quantum system. \textbf{(a)}~Input quantum circuit and hardware topology, consisting of two chips connected by an inter-chip coupler. \textbf{(b)}~Min-cut partitioning of the qubit interaction graph, \textbf{(c)}~Final coupler-aware mapping, which minimizes expensive \textbf{Inter-chip Operations}.}
    \label{fig:example}
\end{figure}

\subsection{Motivation}

\noindent\textbf{Insight 1: Distinguish Inter- and On-Chip Operations ---}
In coupler-connected modular quantum systems, inter-chip operations differ from on-chip ones in lower fidelity, higher latency, and supported gate sets. For example, with IBM’s Heron R2, on-chip CNOT gates achieve error rates as low as $8\times10^{-4}$ (e.g., 99.92\% fidelity~\cite{ibm_heron_r2}), while inter-chip CNOTs demonstrated on IBM’s \textit{Flamingo} exhibit the best error rates at $3.5\%$ (e.g., 96.5\% fidelity~\cite{ibmQuantumDelivers}), more than 40 times higher error rate than an on-chip CNOT.

Additionally, native gate sets diverge. For example, while SWAPs in monolithic devices require three CNOTs, dedicated coupler links can implement SWAPs directly through analog exchange~\cite{krizan2024swap}. These analog SWAPs, related to iSWAP up to phase corrections, reduce both execution time and error accumulation.

Therefore, compilation for modular systems should be guided by experimentally informed models that capture the high cost of inter-chip operations. To optimize fidelity, the compiler must explicitly distinguish between on-chip and inter-chip gates, utilize native analog operations when supported, and apply noise-aware mapping strategies to mitigate the impact of hardware variability.

\noindent\textbf{Insight 2: Minimizing Inter-Chip Operations through Entanglement-Aware Partitioning ---}

Reducing inter-chip operations is critical in modular quantum systems due to significantly higher error rates and latency from coupler-based links. Modular compilation addresses this by partitioning circuits into subgraphs mapped to individual chips, with each subcircuit compiled independently. However, naive partitioning may unintentionally place strongly entangled qubits on different chips, increasing costly inter-chip communication.

Figure~\ref{fig:example} shows that simple min-cut approaches can place entangled qubits on different chips, leading to increased inter-chip gates. In contrast, entanglement-aware partitioning clusters highly interactive qubits, isolating operations that incur high coupler noise. Our method explicitly models fidelity variation across couplers and prioritizes keeping strongly connected components within the same chip.

Most existing mapping strategies~\cite{sabre2019, hua2023qasmtrans, cheng2024robust, jin2024tetris} assume monolithic hardware and ignore coupler-induced noise. \sol~extends these by co-optimizing circuit cutting and qubit mapping, leveraging topology-aware partitioning to minimize inter-chip operations and enable scalable, high-fidelity modular execution.
\section{System Model and Problem Formulation}
\label{sec:model}

We consider a coupler-connected modular quantum system comprising \( n \) quantum chips \( \mathcal{C} = \{c_1, c_2, \ldots, c_n\} \). Each chip \( c_i \) serves as a quantum processor with three defining features:
\begin{itemize}
    \item \textbf{On-chip topology}: A heavy-hex lattice \( T_i = (V_i, E_i) \), where \( V_i \) denotes physical qubits arranged with 2–3 nearest neighbors to balance connectivity and crosstalk.
    
    \item \textbf{Noise profile}: \( N_i = \{\epsilon_g^{(i)}, \epsilon_r^{(i)}\} \), where \( \epsilon_g^{(i)} \) is the average two-qubit gate error, and \( \epsilon_r^{(i)} \) is the readout error, derived from hardware calibration.

    \item \textbf{Inter-chip couplers}: A set of fixed links \( K \), where each coupler connects \( q_k \in c_i \) and \( q_l \in c_j \) with an associated error rate \( \epsilon_{\text{coupler}}^{(kl)} \), enabling inter-chip operations.
\end{itemize}

The coupler-connected system enables circuit execution across multiple chips with distinct topologies, noise profiles, and limited inter-chip links. We model the modular system as a unified graph \( G_{\text{MC}} = (V_{\text{MC}}, E_{\text{MC}}) \), where \( V_{\text{MC}} \) includes all physical qubits, and \( E_{\text{MC}} \) combines on-chip and inter-chip connections. The system graph retains each chip's heavy-hex topology through on-chip edges. Inter-chip edges, representing couplers like IBM's L-couplers, are weighted by their cumulative noise contributions. \( w_{kl} = \epsilon_{\text{coupler}}^{(kl)} + \epsilon_g^{(k)} + \epsilon_g^{(l)} \), where \( \epsilon_{\text{coupler}}^{(kl)} \) is the error rate of the coupler connecting qubits \( q_k \) and \( q_l \); and \( \epsilon_g^{(k)}, \epsilon_g^{(l)} \) are gate error rates of qubits \( q_k \) and \( q_l \).

The \( G_{\text{MC}}\) embeds hardware-specific metrics, including SWAP/coupler latencies and qubit decoherence rates (\( T_{1,q}, T_{2,q} \)), to enable noise-aware optimization. This unified representation allows compilers to holistically partition circuits, minimize inter-chip operations, and balance runtime against decoherence limits. Based on MC, our cost function quantifies the trade-offs between operation overhead, runtime, and fidelity degradation when compiling circuits across the \( G_{\text{MC}}\). 

The total cost function, \( C_{total} \), integrates four interdependent components derived from hardware-aware metrics: 

\begin{equation}
    \begin{aligned}
        C_{\text{total}} &= \underbrace{\alpha \cdot S_{\text{on}} + \beta \cdot S_{\text{inter}}}_{\text{Operational Overhead}}\quad + \underbrace{\gamma \cdot D}_{\text{Temporal Cost}} + \underbrace{\delta \cdot \frac{\sum \epsilon}{\Gamma_{\text{avg}}}}_{\text{Fidelity Penalty}}
    \end{aligned}
    \label{equ:2}
\end{equation}

This cost function contains three parts: Operational Overhead, Temporal Cost, and Fidelity Penalty.  
\textbf{Operational Overhead} captures the latency and resource costs of adapting the circuit to the modular system. It combines on-chip SWAP operations and inter-chip coupler usage. The terms \( S_{\text{on}} \) and \( S_{\text{inter}} \) are defined as: $S_{\text{on}} = \sum_{t=1}^{D_{\max}} \sum_{(q_i,q_j) \in E_i} s_{q_i,q_j}^t$ and $ S_{\text{inter}} = \sum_{t=1}^{D_{\max}} \sum_{(q_k,q_l) \in E_{\text{coupler}}} r_{q_k,q_l}^t$

where: \( s_{q_i,q_j}^t \in \{0,1\} \) is binary variable indicating a SWAP operation between qubits \( q_i \) and \( q_j \) at timestep \( t \); \( r_{q_k,q_l}^t \in \{0,1\} \): Binary variable indicating an inter-chip coupler operation between \( q_k \) and \( q_l \) at timestep \( t \); and \( D_{\max} \): Maximum circuit depth (total timesteps).

The \textbf{Temporal Cost \( D \)} measures total execution time, factoring in the slowest chip operations and delays introduced by inter-chip couplers:
\begin{equation}
    D = \underbrace{\max_{c_i} \left( \text{Depth}(c_i) \right) \times t_{\text{layer}} }_{\text{Local Computation}} + \underbrace{t_{\text{coupler}} \cdot S_{\text{inter}}}_{\text{Coupler Delay}},
    \label{equ:5}
\end{equation}
where:  \( \text{Depth}(c_i) \) is Circuit depth (number of layers) on chip \( c_i \); \( t_{\text{layer}} \) is Time per circuit layer (e.g., longest gate time in a specific layer); and \( t_{\text{coupler}} \) is the time delay per coupler operation.

The \textbf{Fidelity Penalty} quantifies cumulative errors from gates, SWAPs, and couplers, normalized by the average decoherence rates of the involved chips:
\begin{equation}
    \text{Fidelity Penalty} = \delta \cdot \frac{\sum \epsilon}{\Gamma_{\text{avg}}},
    \label{equ:6}
\end{equation}
with the total error \( \sum \epsilon \) defined as: $\sum \epsilon = \sum_{g \in G_{\text{on}}} \epsilon_g^{(i)} + \sum_{s \in S_{\text{on-chip}}} 3\epsilon_g^{(i)} + \sum_{r \in S_{\text{inter}}} \left( \epsilon_{\text{coupler}}^{(kl)} + \epsilon_g^{(k)} + \epsilon_g^{(l)} \right)$,

where: \( \sum_{g \in G_{\text{on}}} \epsilon_g^{(i)} \) is total gate errors in on-chip operations, 
\( 3.5\epsilon_g^{(i)} \) is the Error amplification per SWAP,
and \( \epsilon_{\text{coupler}}^{(kl)} + \epsilon_g^{(k)} + \epsilon_g^{(l)} \) is the cumulative error for inter-chip coupler operations.

The normalization factor \( \Gamma_{\text{avg}} \) penalizes fidelity loss based on chip-averaged decoherence:
\begin{equation}
    \Gamma_{\text{avg}} = \frac{1}{n} \sum_{c_i \in \mathcal{C}} \left( \frac{1}{T_{1,\text{avg}}^{(i)}} + \frac{1}{T_{2,\text{avg}}^{(i)}} \right),
    \label{equ:8}
\end{equation}
where: $T_{1,\text{avg}}^{(i)}$, $T_{2,\text{avg}}^{(i)}$ are average $T_1, T_2$ on the chip $i$ and $n$ is the total number of chips in \( \mathcal{C} \).

The information required by our problem formulation can all be provided by system calibration data, such as per-qubit error rate, gate time and $T_1, T_2$. Based on the problem formulation, we aim to find a mapping that minimizes the cost.
\section{\sol~Compilation Framework}

\begin{figure}[htbp]
    \centering
     \includegraphics[width=1\linewidth]{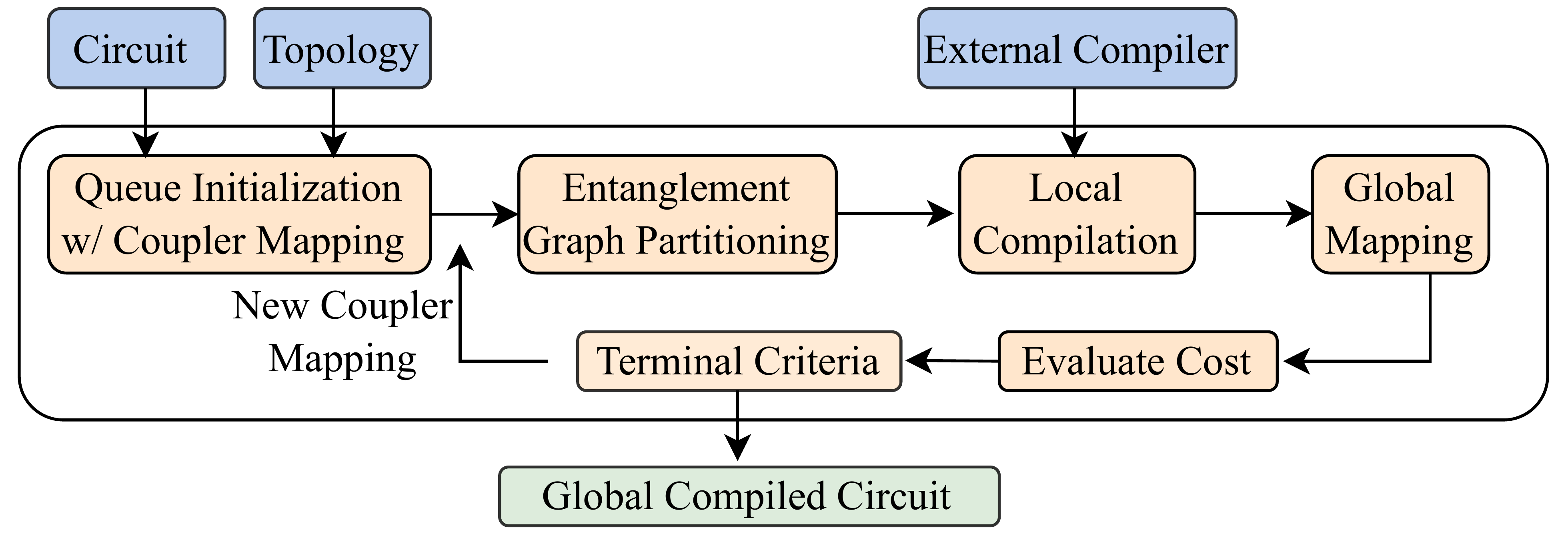}
    \caption{
System overview of the \sol~framework. The compilation loop dynamically alternates between local compilation and global mapping, guided by cost evaluation and coupler-aware optimization.
    }
    \label{fig:overview}
\end{figure}

\begin{figure}[htbp]
    \centering
    \includegraphics[width=1\linewidth]{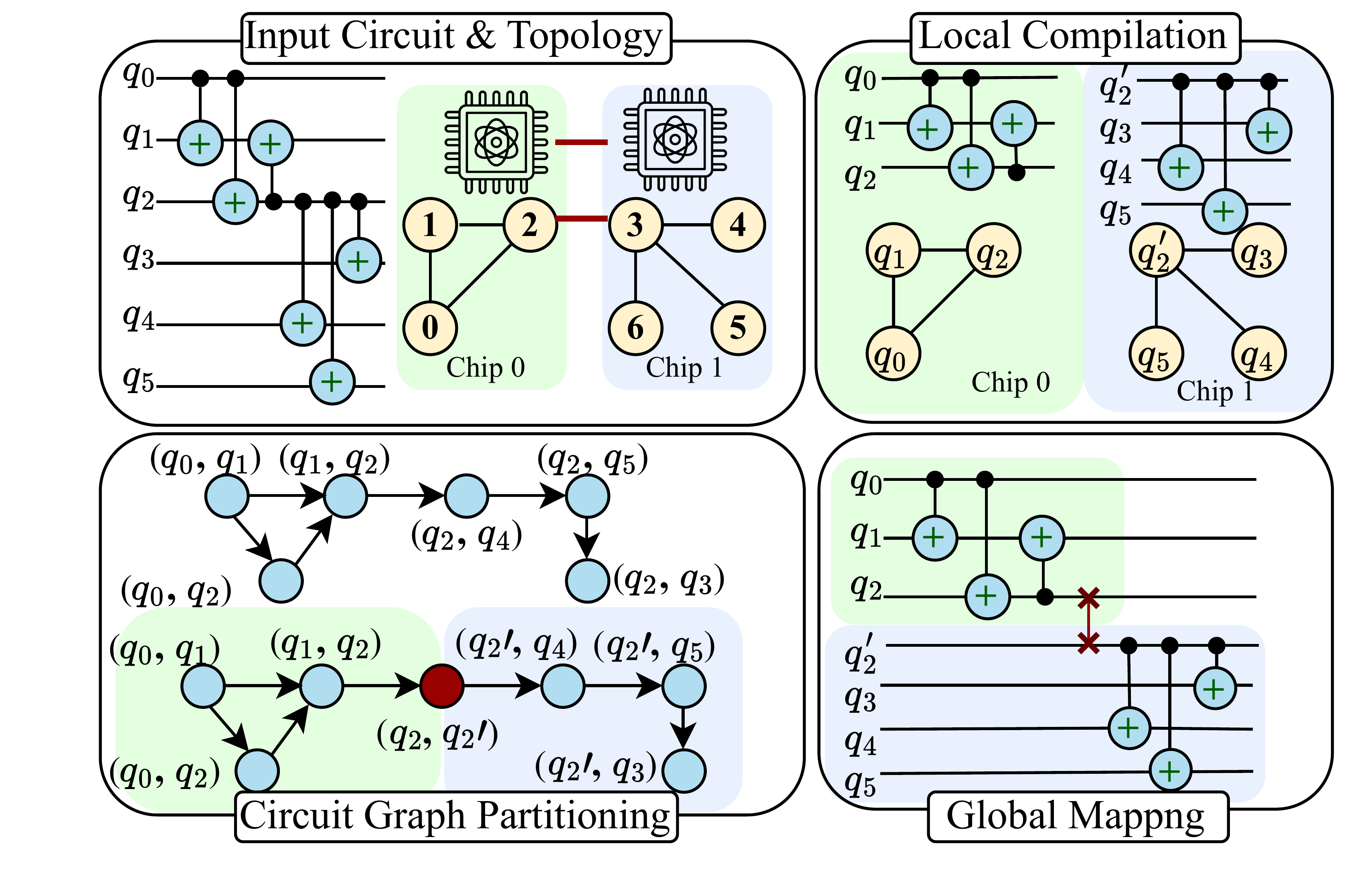}
    \caption{
    \sol~Workflow illustrates on a two-chip system, where CCMap’s partitioning avoids inter-chip operations. Given a circuit and hardware topology, \sol~ performs entanglement-based partitioning followed by local compilation. Finally, it applies global mapping to avoid inter-chip operations and produces a topology-aware output
    }
    \label{fig:process}
\end{figure}

Figure~\ref{fig:overview} shows the overall \sol~pipeline, while Figure~\ref{fig:process} illustrates its operation on a small two-chip example. \sol~targets modular quantum systems with fixed interconnects and heterogeneous noise. The compilation begins by analyzing the circuit’s entanglement graph to partition it into fragments. Fragments are determined by selecting physical chips with optimal decoherence performance and setting a target number of fragments based on available chip capacity.

Each fragment is locally compiled using external compilers, and~\sol~subsequently assembles these into a global circuit by inserting inter-chip operations along fragment boundaries. A priority queue is maintained to track available physical chips, and candidate mappings are greedily refined using the total cost that captures the overall noise level of the assembled circuit. This dynamic, calibration-driven approach enables~\sol~to yield high-fidelity global circuit mappings, even with local search strategies.

\subsection{Input Data and External Compiler}

Our framework accepts three inputs: (i) the quantum circuit to be compiled, (ii) an external compiler used for local compilation, and (iii) the hardware topology as a weighted graph\(G(V,E)\), where each vertex \(v \in V\) represents a physical qubit and each edge \(e(v_i,v_j) \in E\) represents a connection between two qubits. The topology is augmented with the latest calibration data from the chip-to-chip connected system. In particular, each qubit is characterized by:
$ 
\{T_1(v),\,T_2(v),\,\epsilon_g(v)\},
$
where \(T_1(v)\) is the qubit's relaxation time, \(T_2(v)\) its dephasing time, and \(\epsilon_g(v)\) its single-qubit gate error rate. These parameters are incorporated into our total cost function to inform inter-chip operation cost.

For any on-chip connection, the edge weight is set equal to the two-qubit gate error. In contrast, for an inter-chip connection, the edge weight is defined by the coupler-specific error rate for that link, which accounts for the distinct noise characteristics of supported operations. Since inter-chip connections incur higher error rates, they are assigned larger weights, guiding the compilation algorithms to minimize costly inter-chip operations.

\sol~adaptively incorporates the most recent calibration data in each compilation cycle. By leveraging real-time calibration data, this approach keeps compilation decisions in sync with the system’s current hardware state, thereby improving both fidelity and execution efficiency

\subsection{Two-Phased Compilation}  \label{subsec:Optimization_Overview}

Algorithm~\ref{alg:multichipmapping} outlines the \sol~pipeline, which uses a two-phased local-global strategy to compile circuits onto coupler-connected modular quantum systems.

The process begins by \textbf{ranking} each chip’s average decoherence rate, \(\Gamma_{\text{avg}}^{(j)}\), based on calibration data such as \(T_1\) and \(T_2\) times (lines 1–3; Equation~\ref{equ:8}). Using this information and the total number of logical qubits from each chip capacity, the algorithm determines the number of circuit partitions \(k\) and assigns a subset of chips with the best decoherence profiles of coupler mapping (lines 4–5).

A priority queue \(Q\) is then initialized with a set of candidate mappings and associated costs. The iterative optimization process begins by selecting a candidate mapping \(M_G\) from the queue (line 6-9). The circuit’s \textbf{entanglement graph} \(G_{qc}\) is partitioned using a community clustering algorithm~\cite{kan2024scalable} to produce \(k\) logical subcircuits \([f_i]_k\), minimizing qubit interactions that span chips (line 10).

Each logical fragment is \textbf{locally compiled} on its assigned chip using an external compiler (line 11), with physical mapping constrained by inter-chip coupler boundaries. The compiled subcircuits are merged in a \textbf{global mapping} step and produce a full circuit \(M_{qc}\) (line 12). This global circuit is evaluated using the noise-aware cost function defined in Equation~\ref{equ:2} (line 13).

The best-known result is updated if the current mapping yields a lower cost than previously found solutions (lines 14–16). New neighbor mappings are then generated by exploring alternative chip-coupler pairings and appended back to the queue (lines 17–18). The algorithm repeats this process until the queue is exhausted or the maximum number of iterations \(I_{\max}\) is reached (lines 19–20). The final output is a \textbf{coupler-aligned} globally compiled, coupler-aligned quantum circuit optimized for modular execution.

\begin{algorithm}[H]
 \caption{Two-phased Compiler Optimization}
\begin{algorithmic}[1]
\REQUIRE 
Circuit $qc$, Entangling Graph $G_{qc}$ 
Chips($[G_i,E]$), 
External Compiler, Max Iteration $I_{\max}$
\ENSURE Mapping $M_{qc}(q\rightarrow Q)$

\FOR{$G_i$ in $[G_i,E]$}
    \STATE G.decoherence $ \Gamma_{\text{G}}$ $\gets$ Equation~\ref{equ:8}
\ENDFOR
\STATE k $ \gets$  QC, $[G_i,E]$ 
\STATE $[G_i,E_i]_k \gets $ Sort $G_i$ by decoherence
\STATE Priority Queue $Q$ Initialization: \{$M_G:[G_i]_k \rightarrow \emptyset$, cost: $\infty$\}
\STATE$M_{qc}$ $\gets$ null, bestCost $\gets \infty$, iteration $\gets 0$
\WHILE{$Q \neq \emptyset$ \textbf{and} iteration $< I_{\max}$}
    \STATE $M_G$ $\gets Q.\text{pop()}$
    \STATE $[f_{i}]_k$, constraint $\gets$ community\_clustering($G_{qc}$, $k$)
    \STATE $M_G:[G_i]_k\rightarrow[f_i]_k$ $\gets$ External Compiler ($[f_i]_k,[G_i]_k$,constraint)
    \STATE $M_{qc}$ $\gets$ global mapping($M_G, [G_i,E_i]_k$)
    \STATE $\text{C}_{now}\gets$ Cost($M_{qc}(q\rightarrow Q)$)
    \COMMENT{See Equation~\ref{equ:2}}
    \IF{$\text{C}_{now}  < \text{C}_{best}$}
         \STATE $\text{C}_{best}, M_{qc} \gets \text{C}_{now}, M_{qc}$ 

    \ENDIF
    \STATE $[M_{G}] \gets$ generate neighbors($M_G$)
    \STATE $Q.\text{append}$($[M_{G}]$)
    \STATE iteration $\gets$ iteration + 1
\ENDWHILE
\RETURN \(M_{qc}\)
\end{algorithmic}
\label{alg:multichipmapping}
\end{algorithm}

\begin{figure*}[htbp]
    \centering
    \includegraphics[width=1\linewidth]{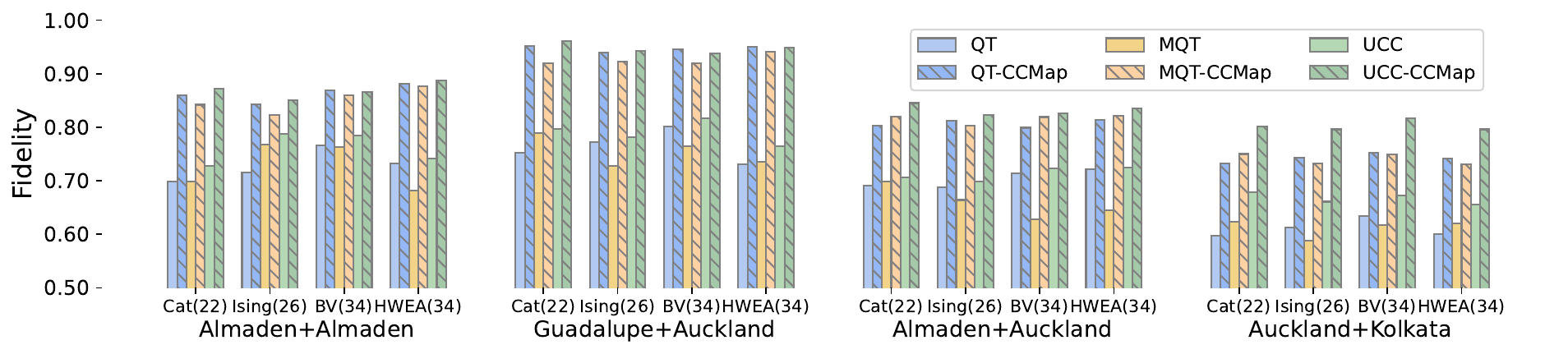}
    \caption{Fidelity on 1-Link Connected Modular Quantum System.}
    \label{fig:Fidelity}
\end{figure*}

\subsection{\sol~Modules}

This section describes the three main modules in Algorithm~\ref{alg:multichipmapping}: entangling graph partitioning, constrained local compilation, and global mapping. Each addresses key stages of compiling circuits for coupler-connected modular quantum systems.

\subsubsection{Entangling Graph Partitioning}

Given a circuit \(C\), we construct an entangling graph where nodes represent two-qubit gates and edges reflect their connectivity. The circuit is partitioned into \(k\) fragments, based on logical qubit count and per-chip capacity, such that inter-fragment (cut) edges are minimized. Our partitioning algorithm, inspired by community detection heuristics~\cite{kan2024scalable}, preferentially groups logical qubits with low interaction weights while cutting high-cost edges. The output is a set \(\{s_i\}\) of fragments mapped to physical chips with minimal inter-chip operation overhead.

\subsubsection{Constrained Local Compilation}

After partitioning, each fragment \(s_i\) is mapped onto its designated chip \(c_j\) using an external compiler. This local compilation phase is performed on a per-chip and per-fragment basis using either swap-count-based or noise-aware techniques. Because of hardware constraints, only physical qubits connected to an inter-chip coupler can be swapped between chips. To preserve the circuit order, additional mapping constraints are enforced. Specifically, any qubits crossing fragment boundaries are pre-mapped to the physical qubit linked to the inter-chip coupler at the end of one fragment and similarly assigned in the subsequent fragment. The associated cost (e.g., additional SWAPs) is incorporated into the overall compilation cost.

\subsubsection{Global Mapping}

Once all fragments are locally compiled, the individual fragments are assembled into a global circuit. Inter-chip operations and cross-fragment scheduling are inserted where needed to maintain logical dependencies. The assembled global circuit is then evaluated using our cost function (Equation~\ref{equ:2}), which considers gate errors, SWAPs, coupler usage, and decoherence. The final output is a low-cost, coupler-aware, minimum-cost, and tailored compiled circuit for the chip-to-chip modular system.
\section{Performance Evaluation}

\begin{table*}[ht]
\centering
\caption{Inter-Chip and On-Chip Operations with 1-link System}
\resizebox{1\textwidth}{!}{%
\begin{tabular}{|c|c|ccc|ccc|ccc|ccc|ccc|ccc|}
\hline
\multirow{2}{*}{\textbf{System}} & \multirow{2}{*}{\textbf{Circuit (size)}}
& \multicolumn{3}{c|}{\textbf{QT}} 
& \multicolumn{3}{c|}{\textbf{QT-CCMap}} 
& \multicolumn{3}{c|}{\textbf{MQT}} 
& \multicolumn{3}{c|}{\textbf{MQT-CCMap}}
& \multicolumn{3}{c|}{\textbf{UCC}} 
& \multicolumn{3}{c|}{\textbf{UCC-CCMap}} \\ 
\cline{3-20}
& & 
\makecell{Inter} & \makecell{On} & \makecell{Depth} & \makecell{Inter} & \makecell{On} & \makecell{Depth} &\makecell{Inter} & \makecell{On} & \makecell{Depth} &\makecell{Inter} & \makecell{On} & \makecell{Depth} &\makecell{Inter} & \makecell{On} & \makecell{Depth} &\makecell{Inter} & \makecell{On} & \makecell{Depth} 
\\
\hline

\multirow{4}{*}{\makecell{AlmadenV2(20) \\ $\times$2}} 
& Cat (35) 
& 1 & 0 & 38
& 1 & 0 & 24 
& 1 & 3 & 43
& 1 & 2 & 23 
& 1 & 0 & 41 
& 1 & 0 & 24  \\
& Ising (34) 
& 2 & 0 & 10 
& 1 & 0 & 8 
& 2 & 20 & 29 
& 1 & 23 & 43
& 2 & 0 & 21
& 1 & 0 & 20 \\
& W-State (36) 
& 2 & 0 & 73 
& 2 & 5 & 44
& 2 & 19 & 88 
& 2 & 23 & 60
& 2 & 0 & 144
& 2 & 0 & 78 \\
& GHZ (40) 
& 3 & 21 & 61 
& 1 & 5 & 23 
& 3 & 7 & 51 
& 1 & 0 & 16 
& 2 & 5 & 46 
& 1 & 3 & 22 \\
\hline
\multirow{4}{*}{\makecell{Auckland(27)\\ $\times$3}} 
& Cat (65) 
& 2 & 33 & 98 
& 2 & 15 & 39 
& 2 & 12 & 83 
& 2 & 9 & 30 
& 2 & 23 & 71 
& 2 & 17 & 31 \\
& Ising (66) 
& 4 & 28 & 18 
& 2 & 22 & 18 
& 4 & 80 & 36 
& 2 & 123 & 63 
& 4 & 26 & 21 
& 2 & 27 & 21 \\
& W-State (76) 
& 6 & 91 & 199 
& 4 & 25 & 57
& 6 & 48 & 171 
& 4 & 60 & 75 
& 4 & 34 & 304 
& 2 & 38 & 98 \\
& GHZ (78) 
& 3 & 24 & 120 
& 2 & 34 & 37 
& 3 & 15 & 100 
& 2 & 9 & 30 
& 2 & 16 & 84
& 2 & 19 & 27   \\
\hline
\multirow{4}{*}{\makecell{AlmadenV2(20)$\times$2 \\ + \\ Auckland(27)$\times$2}}
& Cat (65) 
& 3 & 0  & 66 
& 2 & 23 & 40  
& 3 & 9  & 83 
& 2 & 9  & 30  
& 2 & 0  & 71
& 2 & 0  & 31  \\ 
& Ising (66) 
& 6 & 0  & 9
& 2 & 23 & 17
& 6 & 52 & 36 
& 2 & 111 & 51
& 4 & 0  & 21 
& 2 & 0  & 21  \\ 
& W-State (76) 
& 6 & 98  & 198
& 4 & 24  & 57 
& 6 & 45  & 175
& 4 & 55  & 69
& 6 & 21  & 304 
& 4 & 23  & 98   \\ 
& GHZ (78) 
& 3 & 36  & 113
& 2 & 26  & 39
& 3 & 12  & 100  
& 2 & 14  & 28
& 3 & 0   & 84 
& 2 & 0  & 27  \\
\hline
\end{tabular}%
}
\label{tab:1-link}
\vspace{-0.1in}
\end{table*}

\begin{table}[ht]
\centering
\caption{Transpilation Runtime (in Seconds)}
\label{tab:runtime}
\resizebox{0.48\textwidth}{!}{%
\begin{tabular}{|c|c|cc|cc|cc|}
\hline
\textbf{System} & \textbf{Circuit (size)} & {\makecell{QT}} & {\makecell{QT-\\CCMap}} & {\makecell{MQT}}& {\makecell{MQT-\\CCMap}} &
{\makecell{UCC}} & {\makecell{UCC-\\CCMap}}  \\
\hline
\multirow{4}{*}{\makecell{Auckland(27) \\ $\times$3}} 
& Cat (65)   & 0.08 & 0.09 & 0.06 & 0.10 & 0.07 & 0.02 \\
& Ising (66) & 0.22 & 0.48 & 0.47 & 0.64 & 0.02 & 0.03\\
& W-State (76) & 0.13 & 0.34 & 0.18 & 0.72 &  0.03 & 0.03\\
& GHZ (78)   & 0.10 & 0.13 & 0.08 & 0.14 & 0.05 & 0.02 \\
\hline
\multirow{4}{*}{\makecell{AlmadenV2(20)$\times$2 \\ + \\ Auckland(27)$\times$2}} 
& Cat (65)   & 0.78 & 0.13 & 0.11 & 0.13 & 0.07 & 0.03 \\
& Ising (66) & 1.10 & 0.55 & 0.47 & 0.64 & 0.03 & 0.04\\
& W-State (76) & 1.35 & 0.34 & 0.18 & 0.43 & 0.03 & 0.03 \\
& GHZ (78)   & 0.15 & 0.13 & 0.24 & 0.15 & 0.02 & 0.02 \\
\hline
\end{tabular}
}
\end{table}

In this section, we evaluate the performance of \sol~ with a focus on fidelity, inter- and on-chip operations, integrated cost (from Equation~\ref{equ:2}), and runtime with varied modular chips. 

\subsection{Implementation and Workloads}

\sol~ is implemented in Python 3.10 using Qiskit 1.2, and tested on a Google Cloud e2-highmem-16 instance with AMD Rome x86/64 processors. We benchmark a diverse set of circuits with number of qubits (22 to 420). These include widely used quantum algorithms and subroutines, as well as established benchmark suites from related studies\cite{jeng2025modularcompilationquantumchiplet,sabre2019}. 
The 7 types of circuits are \textbf{BV:} Bernstein--Vazirani algorithm for hidden bit string determination~\cite{bernstein1993quantum}. \textbf{Adder:} Quantum ripple-carry adder circuit for quantum arithmetic~\cite{zhang2023characterizing}.  \textbf{HWEA:} Hardware Efficient Ansatz used in variational algorithms~\cite{kandala2017hardware}.  \textbf{ISING:} Ising model circuit for quantum simulation~\cite{lucas2014ising}. \textbf{CAT:} Schr{"o}dinger cat state preparation~\cite{mirrahimi2014cat}. \textbf{W-State:} Circuit for generating robust symmetric entanglement among qubits~\cite{cross2017open}. \textbf{GHZ:} GHZ state preparation for testing nonlocality and quantum correlations~\cite{greenberger1989going}.

Notably, \sol~ works on top of existing quantum compilers and the community detection graph partitioning inspired by Fitcut~\cite{kan2024scalable}. We compare {\bf X-CCMap} against three existing compilers ({\bf X}) under the chip-to-chip coupler-connected modular system: (1) {\bf Qiskit-Transpiler (QT)}: IBM-Q’s default quantum compiler for monolithic devices~\cite{sabre2019}; (2) {\bf MQT}: a framework for efficient noise-aware qubit mapping~\cite{mqtqmap} from Munich Quantum Toolkit; (3) {\bf UCC}: a lightweight and extensible compiler framework for quantum circuit transformation and optimization, developed as part of the Unitary Compiler Collection~\cite{ucc2025}. Therefore, our solution is denoted as {\bf QT-CCMap}, {\bf MQT-CCMap} and {\bf UCC-CCMap}.

\subsection{Evaluation Settings}
We evaluated coupler-connected systems comprising 2 to 4 chips (with 20, 27, or 127 qubits per chip). Due to the limited availability of physical chip-to-chip systems, our experiments utilized IBM-Q noisy emulators~\cite{qiskit_fake_provider} calibrated using real quantum devices (e.g., AlmadenV2, Auckland, and WashingtonV2). The noise profiles, specifically, the parameters as described in Section~\ref{sec:model}, are obtained via the emulator’s API.

For our cost function, Equation~\eqref{equ:2}, we set the weight factors \(\alpha\) and \(\gamma\) to 1 to balance the operational overhead and temporal cost. Additionally, we set the inter-chip coupler error rate \(\epsilon_g^{(kl)}\) to 3.5\% with \(\beta = 3.5 \times \alpha\), following IBM's technical report on IBM's L-couplers~\cite{ibmQuantumDelivers} and  the coupler delay to 30~ns \(t_{\text{coupler}}\) in Equation~\ref{equ:5}. Finally, the depth penalty factor is set as \(\delta = \max_{c_i} \left(\text{Depth}(c_i)\right)\), reflecting the accumulation of noise with increased circuit depth.

\subsection{Evaluation Metrics}

We primarily focus on the following evaluation metrics.
\textbf{Fidelity:} Measures how closely the executed circuit’s final state approximates the ideal outcome by aggregating errors from gates, SWAPs, and couplers. 
 \textbf{Inter-chip Operations:} Quantifies the extra time and error cost from inter-chip (CX, SWAP and State Transfer) operations performed within a coupler across chips.
 \textbf{On-chip Operations:}  the total number of gates executed within individual chips versus those executed across chips via couplers.
\textbf{Total Cost:} As shown in Equation~\ref{equ:2}, it combines the time-equivalent overheads of on-chip and inter-chip operations, additional circuit depth, and fidelity penalties into a single unified metric.
\textbf{Runtime:} Evaluates the overall execution time of the transpiled circuit.
\textbf{Depth:} Refers to the longest sequence of quantum operations that must be executed sequentially.

\subsection{Experiment Results}

Existing chip-to-chip modular quantum systems, such as IBM-Q Flamingo and Crossbill, typically links identical chips. We extend them to diverse architectures, connecting chips with various topology setups: \textbf{Homogeneous System}: Multiple identical chips are connected with couplers, e.g., AlmadenV2(20)$\times2$ means two AlmadenV2(20) chips are connected by couplers for a new modular system with 40 qubits.
\textbf{Heterogeneous System}: multiple different chips that have diverse error rate, qubit capacities and topologies are connected, e.g., AlmadenV2(20) $\times 2 +$ Aukland(27) $\times 2$, the system has a total 94 qubit capacity.

\subsubsection{Fidelity}

Figure~\ref{fig:Fidelity} presents the fidelity across 4 settings with different chips, including Almaden, Guadalupe, Auckland, and Kolkata, which offer qubit capacities of 20, 16, 27, and 27, respectively. Obviously, \sol~ optimized compilers consistently achieve the highest fidelity in all scenarios than baseline methods (external compilers) across various modular configurations. Since \sol~can enhance fidelity by partitioning the circuit and exploiting cross-chip flexibility. Based on the cost function, Equation 1,~\sol's qubit placement avoids low-quality physical qubits on each chip and high-cost inter-chip operations on coupler cross chips. On average specifically, QT-CCMap improves fidelity from 70.2\% to 84.0\% (a 13.8\% improvement and 19.7\% increase from QT), MQT-CCMap raises it from 68.8\% to 83.3\% (a 14.5\% improvement and 21.1\% increase from MQT), and UCC-CCMap enhances it from 73.3\% to 86.3\% (a 13.0\% improvement and 17.8\% increase from UCC). These consistent gains across all compilers confirm that \sol's coupler-aligned and hardware-aware mapping and partitioning strategies are critical for preserving fidelity in noisy modular architectures.

Specifically, in Cat(22) executed on an Guadalupe(16) + Auckland(27) system, UCC achieves a fidelity of 80.0\% with 1 inter-chip operation, while \sol~ optimized solution, UCC-CCMap, reaches 96.1\% with 0 inter-chip operation, which is a 16.1\% improvement, 20.1\% increase from original UCC. Similarly, for the HWEA(34) circuit under the same hardware configuration, the fidelity improves from 73.1\% with 3 inter-chip operations (QT) to 95.0\% with 0 inter-chip operations (QT-CCMap), representing a 21.9\% increase, 30.0\% improvement from the original compiler. The decreasing inter-chip operations by \sol~leads to the fidelity increment.

Across other benchmarks, such as the Cat(22) circuit on homogenouse two AlmadenV2(20) system, QT-CCMap achieves a fidelity of 86.0\%, while MQT-CCMap achieves 84.2\%, both higher than their baselines (QT: 69.8\%, MQT: 70.1\%). In the Ising(26) benchmark on the heterogeneous system with 2 chips GuadalupeV2(16)+Auckland(27), QT-CCMap reaches 84.3\% versus 71.6\% from QT. These results further demonstrate that while variations in chip-specific errors impact fidelity for the same circuit, \sol~consistently achieves significant fidelity improvements across different system configurations.

The trend remains consistent across all modular chip-to-chip connected system configurations: CCMap-optimized compilers consistently outperform their corresponding baselines, QT, MQT and UCC, and achieve significant fidelity improvements. 

Traditional compilers failed to capture the distinct noises from individual chips and coherence behavior associated with coupler connections. This is a significant bottleneck when using compilers designed for monolithic chips. In contrast, our two-phased local-global framework via~\sol~enhances these compilers by focusing on single-chip compilation while strategically minimizing inter-chip operations. Moreover, by integrating a cost function that accounts for the global error accumulation over time,~\sol~achieves higher fidelity without being constrained by compiling the partitions locally.

\begin{table}[ht]
\centering
\caption{Total Cost Comparison ($\times 10^{-3}$ for all entries)}
\resizebox{0.5\textwidth}{!}{%
\begin{tabular}{|c|c|cc|cc|cc|}
\hline
\textbf{System} & \textbf{Circuit} & {\makecell{QT}} & {\makecell{QT-\\CCMap}} & {\makecell{MQT}}& {\makecell{MQT-\\CCMap}} &
{\makecell{UCC}} & {\makecell{UCC-\\CCMap}} \\
\hline
\multirow{4}{*}{\makecell{AlmadenV2(20) \\ $\times$2}}
& Cat (35) 
& 2.1 & 1.4 & 3.0 & 2.2 & 1.4 & 0.9 \\
& Ising (34) 
& 0.6 & 0.4 & 8.0 & 4.3 & 0.7 & 0.6 \\
& W-State (36) 
& 4.4 & 2.9 & 8.1 & 6.8 & 5.2 & 3.0 \\
& GHZ (40) 
& 5.8 & 2.4 & 4.9 & 3.2 & 1.8 & 0.9 \\
\hline
\multirow{4}{*}{\makecell{Auckland(27) \\ $\times$3}}
& Cat (65) 
& 13.5 & 8.8 & 8.8 & 10.1 & 4.6 & 2.0 \\
& Ising (66) 
& 5.8 & 5.7 & 15.6 & 33.6 & 1.4 & 1.4 \\
& W-State (76) 
& 34.2 & 22.0 & 39.6 & 23.2 & 23.1 & 8.0 \\
& GHZ (78) 
& 17.1 & 13.9 & 15.6 & 10.7 & 6.6 & 2.2 \\
\hline
\multirow{4}{*}{\makecell{AlmadenV2(20)$\times$2 \\ + \\ Auckland(27)$\times$2}} 
& Cat (65) 
& 6.5 & 5.8 & 9.6 & 8.3 & 4.6 & 2.7 \\
& Ising (66) 
& 5.8 & 5.7 & 33.6 & 15.6 & 1.4 & 1.7 \\
& W-State (76) 
& 34.2 & 22.0 & 39.6 & 23.2 & 23.5 & 8.2 \\
& GHZ (78) 
& 17.1 & 13.9 & 15.6 & 10.7 & 6.8 & 2.1 \\
\hline
\end{tabular}
\label{tab:cost}
}
\end{table}

\begin{table}[ht]
\centering
\caption{Inter- \& On-chip Overhead on GHZ(40) with Varying Link Counts on AlmadenV2(20)$\times2$ System}
\resizebox{0.5\textwidth}{!}{%
\begin{tabular}{|c|cc|cc|cc|cc|cc|cc|}
\hline
\multirow{2}{*}{\textbf{Link}} 
& \multicolumn{2}{c|}{\makecell{QT}} 
& \multicolumn{2}{c|}{\makecell{QT-\\CCMap}}
& \multicolumn{2}{c|}{\makecell{MQT}} 
& \multicolumn{2}{c|}{\makecell{MQT-\\CCMap}}
& \multicolumn{2}{c|}{\makecell{UCC}} 
& \multicolumn{2}{c|}{\makecell{UCC-\\CCMap}}
\\
\cline{2-13}
 & \makecell{Inter} & \makecell{On}
 & \makecell{Inter} & \makecell{On} & \makecell{Inter} & \makecell{On} & \makecell{Inter} & \makecell{On} & \makecell{Inter} & \makecell{On} & \makecell{Inter} & \makecell{On}\\
\hline
2-link  
& 3 & 24 & 1 & 5 & 2  & 9 & 1 & 3 & 2 & 7 & 1 & 8\\
3-link  
& 3 & 21 & 0 & 0 & 2  & 4 & 1 & 1 &2&9&1&7 \\
4-link  
& 2 & 19 & 0 & 5 & 2  & 7 & 1 & 1 &2&5&1&3\\
\hline
\end{tabular}%
\label{tab:links}
}
\end{table}

\begin{table*}[ht]
\centering
\caption{Modular Quantum Systems built on multiple WashingtonV2(127) Chips}
\resizebox{1\textwidth}{!}{%
\begin{tabular}{|c|c|ccc|ccc|ccc|ccc|ccc|ccc|}
\hline
\multirow{2}{*}{\textbf{Chips}} & \multirow{2}{*}{\textbf{Circuit (size)}}
& \multicolumn{3}{c|}{\makecell{QT}} 
& \multicolumn{3}{c|}{\makecell{QT-CCMap}}
& \multicolumn{3}{c|}{\makecell{MQT}} 
& \multicolumn{3}{c|}{\makecell{MQT-CCMap}}
& \multicolumn{3}{c|}{\makecell{UCC}} 
& \multicolumn{3}{c|}{\makecell{UCC-CCMap}} 
\\
\cline{3-20}
& & \makecell{Inter} & \makecell{On} & \makecell{depth} & \makecell{Inter} & \makecell{On} & \makecell{depth}& \makecell{Inter} & \makecell{On} & \makecell{depth}& \makecell{Inter} & \makecell{On} & \makecell{depth}& \makecell{Inter} & \makecell{On} & \makecell{depth}& \makecell{Inter} & \makecell{On} & \makecell{depth}
\\
\hline
\multirow{2}{*}{3}
& Cat(260) 
& 5 & 397 & 618 
& 2 & 371 & 290 
& 2 & 40 & 335 
& 2 & 36 & 160
& 2 & 38 & 266
& 2 & 32 & 126
\\
& GHZ(255) 
& 3 & 443 & 645 
& 2 & 388 & 285 
& 2 & 40 & 330
& 2 & 28 & 160
& 3 & 39 & 261
& 2 & 36 & 124
\\
\hline
\multirow{2}{*}{4} 
& Ising(420) 
& 12 & 722  & 44 
& 3  & 571  & 33 
& 6 & 931 & 60 
& 3 & 1158 & 58
& 6 & 668 & 26 
& 3 & 524 & 21

\\
& W-State(380) 
& 6 & 1000 & 1246 
& 4 & 1111 & 444 
& 6 & 240 & 887
& 4 & 189 & 367
& 6 & 964 & 1520
& 4 & 872 & 498
 \\
\hline
\end{tabular}%
\label{tab:large-circuit}
}
\end{table*}

\noindent \textbf{Inter- \& On-chip Operations:} Fidelity is highly sensitive to compilation quality, particularly the number of on-chip and inter-chip operations. We analyze operational overhead to understand how \sol~reduces inter-chip gates and balances on-chip operations. Fewer inter-chip gates lead to lower errors, while efficient on-chip scheduling ensures compact circuit depth.
Table~\ref{tab:1-link} compares the operation counts for various circuits executed on homogeneous and heterogeneous Chip-to-Chip Coupler-Connected Modular Quantum Systems with 1-link connectivity, across six different compilation strategies.

CCMap-based methods consistently reduce inter-chip operations. For instance, in the Ising(66) circuit mapped to Auckland$\times$3, QT-, MQT-, and UCC-\sol~achieve just 4, 4, and 2 inter-chip operations, respectively, compared to 6, 6, and 4 with QT, MQT, and UCC. This reduction stems from \sol's ability to co-optimize circuit partitioning and qubit mapping based on the modular chip layout.

Furthermore, CCMap variants help lower the circuit depth, defined as \(\delta = \max_{c_i} \left(\text{Depth}(c_i)\right)\), by dividing the circuit into smaller subcircuits \(c_i\) and mapping them independently. In contrast, standard compilers that treat the entire modular system as a monolithic topology often introduce deeper schedules due to suboptimal mapping and excessive routing overhead.

While inter-chip operations are more error-prone and time-consuming than on-chip gates, aggressively minimizing them may increase on-chip overhead. For example, in the Cat(65) on the AlmadenV2$\times$3 + Auckland$\times$3 system, QT-CCMap incurs 23 on-chip operations, compared to 0 with QT.

To balance this trade-off, our cost model (Equation~\ref{equ:2}) integrates both on-chip and inter-chip gate costs, weighted by their respective error rates, execution delays, and contribution to overall circuit depth.

\noindent \textbf{Runtime}: Table~\ref{tab:runtime} presents the runtime corresponding to the experiments in Table~\ref{tab:1-link}. Across all experiments, \sol~optimized compilers demonstrate comparable or even faster compilation.
Notably, in specific cases, such as W-State(76) on the AlmadenV2(20)×2 + Auckland(27)×2 configuration QT-CCMap achieves faster runtime than QT, 0.34 seconds vs. 1.35 seconds. While \sol~introduces overhead in calculating cost function and executing the algorithm, it reduces the problem size , e.g., a large combined topology vs. smaller individual topologies and large input circuits vs. corresponding smaller subcircuits, for the external transpilers.

\noindent \textbf{Total Cost}: We evaluate the total cost using Equation~\ref{equ:2}, which accounts operational overhead, temporal cost and fidelity penalty. Table~\ref{tab:cost} highlights cost reductions across all systems, which is from the experiments in Table~\ref{tab:1-link}. \sol~ significantly lowers total cost by minimizing inter-chip gates and optimizing qubit placement.

Across all benchmarks, incorporating ~\sol into the compilation workflow consistently lowers the total cost.
On the AlmadenV2(20)×2 + Auckland(27)×2 system, the cost for the Ising(66) circuit drops from 0.0396 (MQT) to 0.0156 (MQT-CCMap), resulting in a 53.6\% decrease. From Table~\ref{tab:1-link}, the inter-chip operations drop from 6 to 2 for MQT and MQT-CCMap, respectively. Similar improvements are made in other circuits in different systems. Therefore, this shows that~\sol effectively reduces overhead associated with inter-chip interactions.

These results confirm the importance of inter-chip optimization in modular quantum systems. CCMap enhances external compiler pipelines by minimizing high-error interconnect usage and balancing local operation cost.

\subsubsection{Multi-Link Modular Systems}

We now extend our analysis to more complex scenarios involving 2- to 4-link modular systems to evaluate the scalability and performance of~\sol under increased connectivity.

We evaluate the GHZ(40) circuit on the AlmadenV2(20)$\times$2 system under 2-, 3-, and 4-link settings. Table~\ref{tab:links} reports the inter- and on-chip operations for each case; the 1-link baseline is shown in Table~\ref{tab:1-link}.

Increasing the number of coupler links consistently reduces inter-chip operations. For example, QT reduces from 3 (2-link) to 2 (4-link), while QT-CCMap maintains just 1 inter-chip operation across all configurations, showing strong mapping adaptability. On-chip operations also decrease, e.g., MQT drops from 9 to 7, indicating improved global mapping flexibility with higher connectivity.

These results highlight the performance gains from denser interconnects and support the need for co-design between compiler strategies and hardware topology in modular systems.

\subsubsection{Scalability --- larger topologies}

We assess scalability using large circuits: Ising(420), W-state(380), Cat(260), and GHZ(255) on modular systems composed of WashingtonV2(127)$\times$3 and $\times$4, supporting to 255 and 420 qubits, respectively, in Table~\ref{tab:large-circuit}.

To validate \sol's scalability, we evaluate it on large circuits (e.g, executed across multi-chip systems built from high-capacity devices. These tests confirm that \sol maintains its performance advantage in depth and interconnect reduction as circuit and hardware scale increase.

Across all benchmarks, \sol~demonstrates significant improvements in reducing inter-chip operations, depth, and total cost. For example, in the W-State(380) circuit, QT results in 6 inter-chip operations, while QT-CCMap reduces this to 4. The circuit depth decreases from 1246 to 444, and the total cost drops from 1.102 to 0.519, representing a 52.9\% reduction. Also, UCC-CCMap decreases the inter-chip operations from 6 to 4, and depth from 1520 to 498. Similar patterns are consistently observed across other large-scale circuits, highlighting \sol's effectiveness in leveraging modular topology for efficient, low-overhead compilation.

These results confirm \sol’s scalability and reinforce its value as a compiler framework tailored to large, distributed quantum systems, where minimizing interconnect overhead is critical to achieving high-fidelity.
\section{Conclusion}

In this work, we introduced CCMap, a hardware–software co-designed framework for compiling quantum circuits onto chip-to-chip coupler-connected modular systems. CCMap partitions circuits using entanglement-aware analysis and leverages external monolithic compilers for local compilation, followed by a global mapping step that reduces costly inter-chip operations. By integrating real-time calibration data and a noise-aware cost function, CCMap optimizes mapping and coupler link useage in chip-to-chip modular system. Our evaluations on IBM-Q emulated topologies show that CCMap improves fidelity by up to 21.9\%, representing a 30\% increase, and reduces compilation cost by up to 58.6\% compared to state-of-the-art baselines. These results highlight CCMap’s potential to enable scalable, high-fidelity execution across modular quantum hardware. Future work includes refining the cost model, supporting dynamic coupler configurations, and validating CCMap on emerging quantum platforms.

\bibliographystyle{ieeetr}
\bibliography{refs}

\end{document}